# Magneto-optical-electric joint-measurement scanning imaging system for identification of two-dimensional vdW multiferroic

*Yangliu Wu, Bo Peng**

School of Electronic Science and Engineering, University of Electronic Science and Technology of China, Chengdu 611731, China

*e-mail: bo_peng@uestc.edu.cn

**Sample fabrication**

$NiI_2$ flakes were prepared by a normal and widely-used mechanical exfoliation method. The few-layer $NiI_2$ were exfoliated via PDMS films in a glovebox from $NiI_2$ bulk crystals, synthesized by chemical vapor transport method from elemental precursors with molar ratio Ni:I = 1:2. All exfoliated hBN, $NiI_2$ and graphene flakes were transferred onto pre-patterned Au electrodes on $SiO_2$/Si substrates one by one to create a heterostructure in glovebox and were further in-situ loaded into a microscopy optical cryostat for optical measurements in the glovebox. The whole process of $NiI_2$ sample fabrication and measurement were kept out of atmosphere[1,2].

**Optical measurements by the magneto-optical-electric joint-measurement scanning imaging system (MOEJSI)**

As an advanced imaging system, the magneto-optical-electric joint-measurement scanning imaging system (MOEJSI) brings spectroscopic techniques with unmatched spatial resolution to very low temperature, high magnetic field and high electric field measurements[3]. It was developed for investigating the magnetic and ferroelectric properties and their mutual control through magneto-optical-electric joint-measurements, besides Raman and photoluminescence features[4-7]. In particular, the RMCD loops and imaging, linear dichroism (LD) imaging and polarization-electric field hysteresis loop can be achieved when simultaneously applied high magnetic field (7 T) and electric field (100 V) at low temperature of 10 K.

The MOEJSI system is built based on a Witec alpha 300R Plus low-wavenumber confocal Raman imaging microscope with a spatial resolution reaching diffraction limit, which is integrated with a closed cycle superconducting magnet (7 T) with a room temperature bore and a closed cycle cryogen-free microscopy optical cryostat (10 K) with electronic transport measurement assemblies[8-10], as shown in Fig. 1a. This system achieves multi-field coupling of magnetic fields, electric fields and optical fields and realize high speed, sensitivity and resolution. For use in the room temperature bore superconducting magnet, an extended snout sample mount is specially designed for the cryogen-free microscopy optical cryostat (Fig. 1b). The microscopy optical cryostat directly anchored on the XY scanning stage of Witec Raman system for scanning imaging. The objective mounted in a lens tube and the extended snout enter into the room-temperature bore from top and bottom, respectively (Fig. 1b and 1c).

For RMCD and LD measurements, a free-space 532 nm laser of ~2 μW was linearly polarized at 45° to the photoelastic modulator (PEM) slow axis and sinusoidally phase-modulated by PEM (RMCD: 50 KHz; LD: 100 KHz), with a maximum retardance of λ/4 for RMCD and λ/2 for LD. The light was further reflected by a non-polarizing beamsplitter cube (R/T = 30/70) and then directly focused onto samples by a long working distance 50× objective (Zeiss, WD = 9.1 mm, NA = 0.55). The reflected beam which was collected by the same objective passed through the same non-polarizing beamsplitter cube and was detected by a photomultiplier (PMT). The PMT and PEM coupled with lock-in amplifier and Witec scanning imaging system (Fig. 2 and 3). The RMCD and LD maps as a function of magnetic field are shown in Fig. 4. With a magnetic field, the LD domains change synchronously with the switching of magnetic domains. Therefore, the observed LD signals originate from the magnetic structure, rather than ferroelectric polarization. Only the LD optical method can not identify the existence of ferroelectric polarization and is not reliable. The electric measurements, including *P-E*, *C-E* and *J-E* hysteresis loop must be done and mutual control of magnetic orders and ferroelectric polarization should be observed.

**Data availability**

All data are available from the corresponding author upon reasonable request.

**Acknowledgements** B.P. thanks the financial support from National Science Foundation of China (52021001, 62250073).

**Author contributions** B.P. conceived this work. Y.L.W. and B.P. performed the RMCD and LD experiments. All of authors contributed to the discussion.

**Competing interests** The authors declare no competing interests.

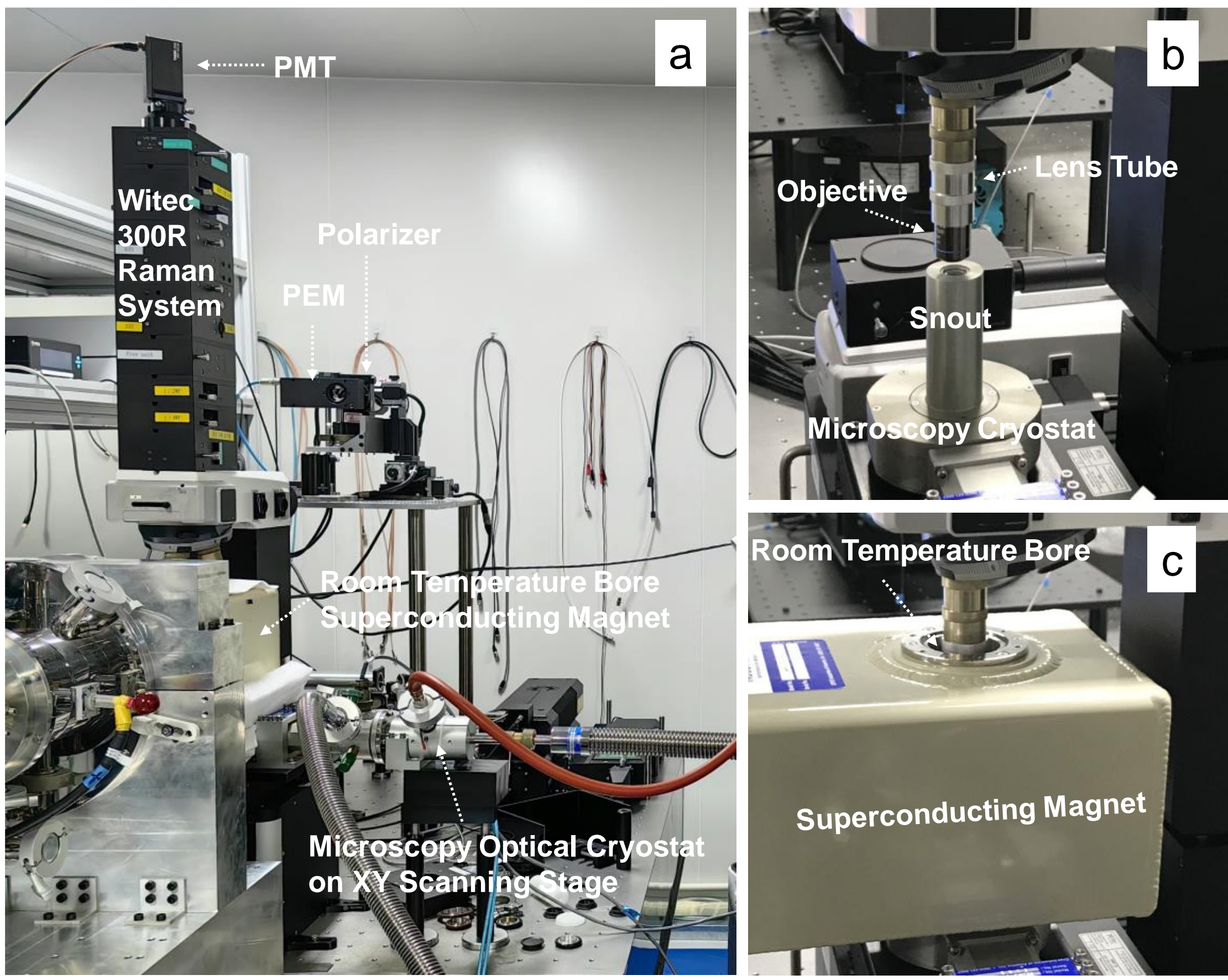

**Fig. 1 | The photograph of the magneto-optical-electric joint-measurement scanning imaging system (MOEJSI). a**, Witec 300R Raman imaging microscope, closed cycle superconducting magnet (7 T) with room-temperature bore and cryogen-free microscopy optical cryostat (10 K) with extended snout sample mount are integrated together. **b**, An objective in an extended tube is coupled with the extended snout of the microscopy optical cryostat. **c**, The objective and the extended snout get into the room-temperature bore of superconducting magnet from top and bottom, respectively.

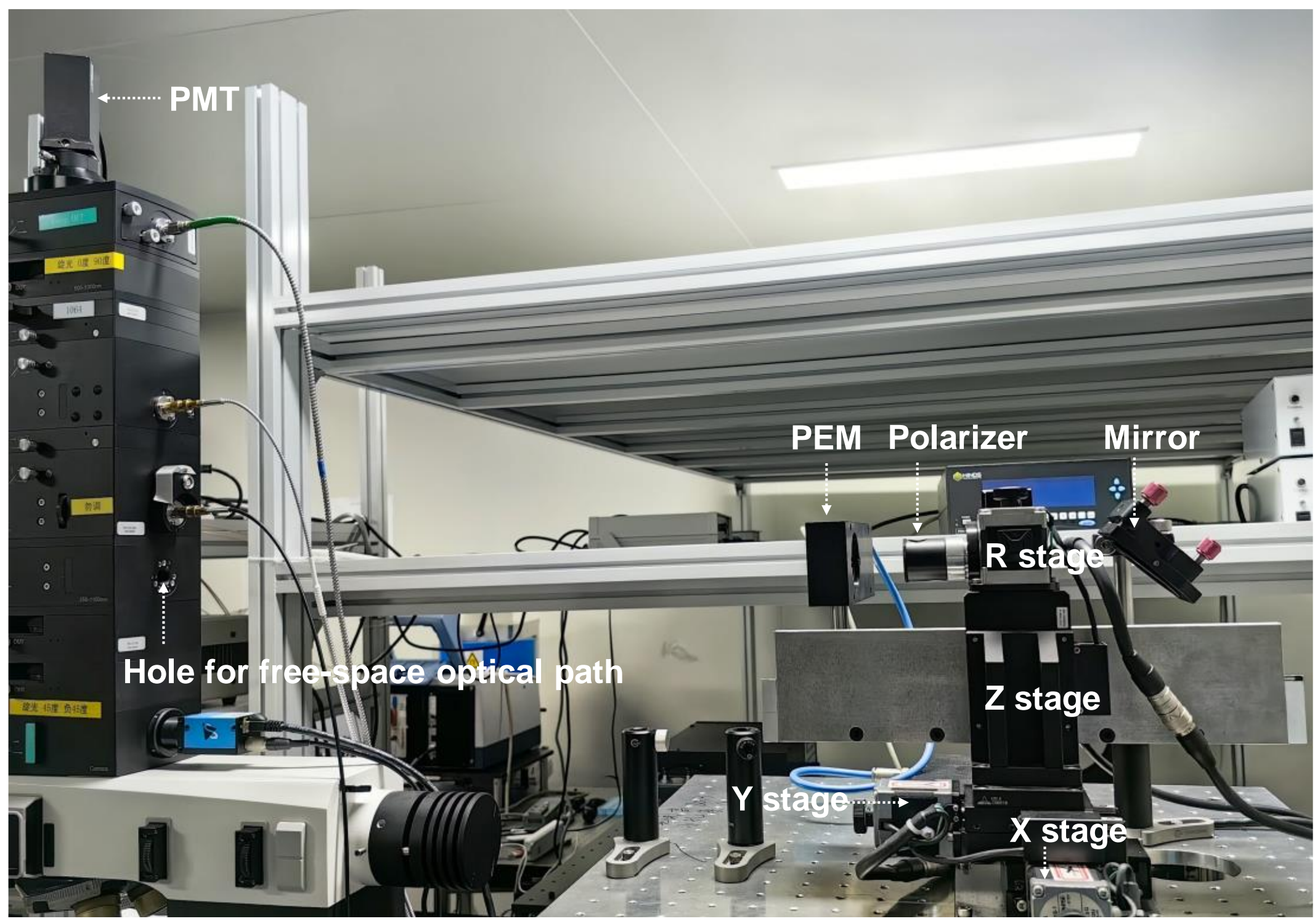

**Fig. 2 | The photograph of free-space RMCD and LD optical path.**

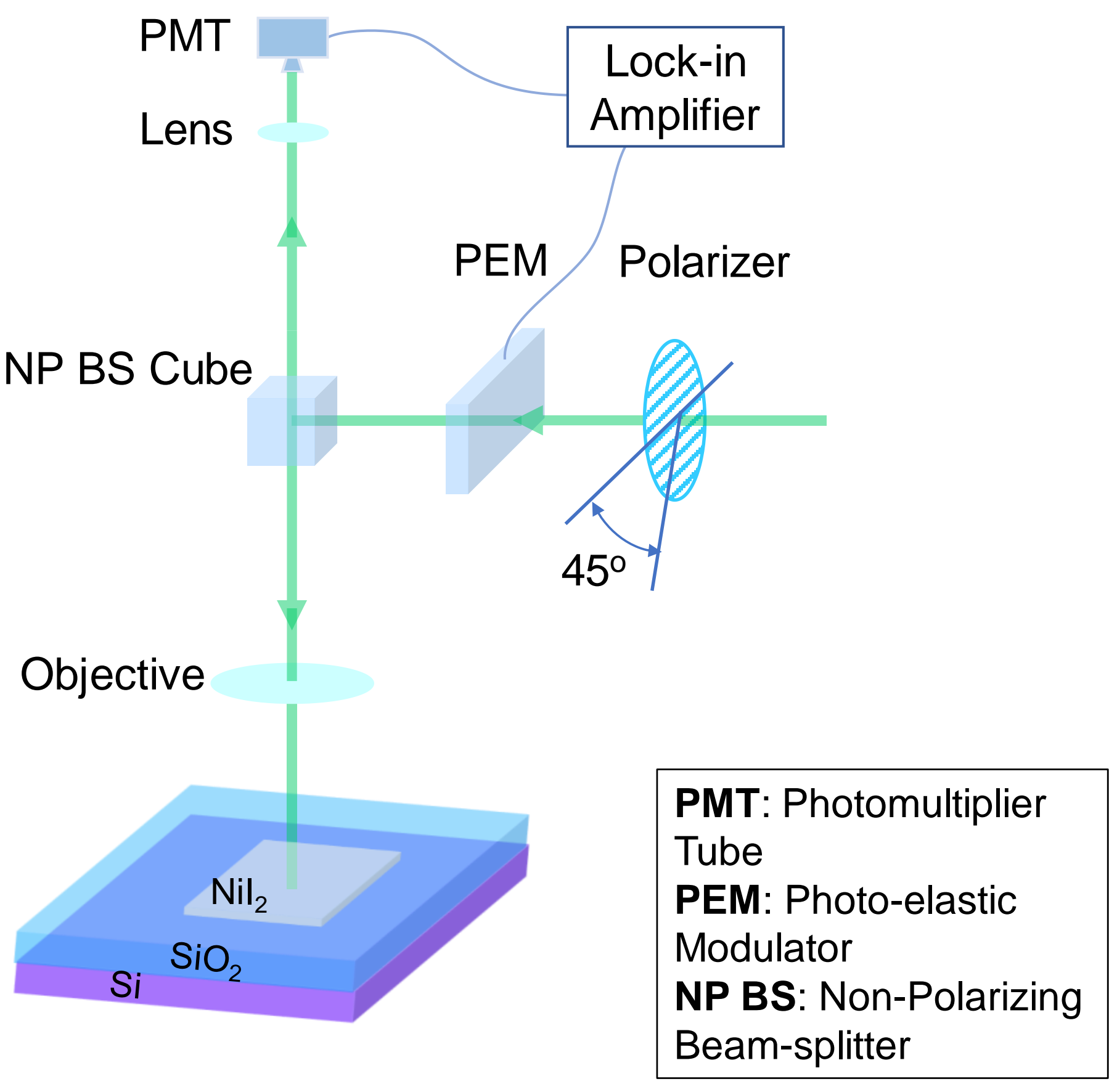

**Fig. 3 | Schematic of free-space RMCD and LD optical path.**

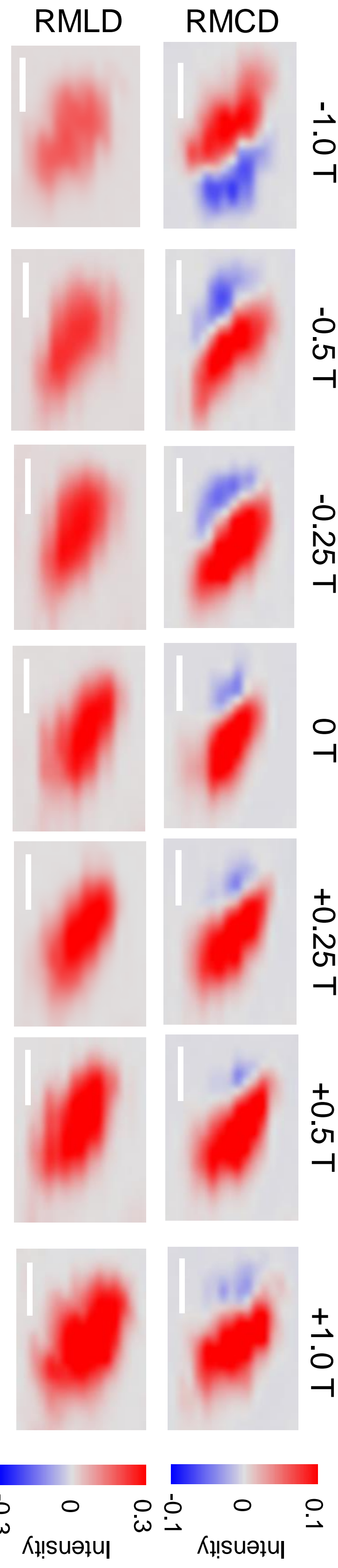

**Fig. 4 | RMCD and LD maps as a function of magnetic field. The scale bar is 1 μm.**